\documentclass{iopart}
\usepackage{iopams,amsopn,graphicx,epsfig,subfigure,times}
\usepackage[usenames]{color}

\renewcommand{\v}[1]{\boldsymbol{#1}}

\newcommand{\h}[2][ ]{\hat{#2}^{\vphantom{\dag} #1}}
\newcommand{\hd}[2][ ]{\hat{#2}^{\dag #1}}

\newcommand{\Var}{\operatorname{Var}}

\newcommand {\Ref}[1] {reference~\cite{#1}}
\newcommand {\Fig}[1] {figure~\ref{#1}}
\newcommand {\Eqn}[1] {equation~(\ref{#1})}

\begin{document}

\title[Multimode analysis of non-classical correlations in double well BECs]{Multimode analysis of non-classical correlations in double well Bose-Einstein condensates}

\author{Andrew~J~Ferris and Matthew~J~Davis}
\address{The University of Queensland, School of Mathematics and Physics, ARC Centre of Excellence for Quantum-Atom Optics, Qld 4072, Australia}
\ead{ferris@physics.uq.edu.au}

\date{\today}
\begin{abstract}
The observation of non-classical correlations arising in interacting two to size weakly coupled Bose-Einstein condensates was recently reported by Est\`eve \emph{et al.} [Nature \textbf{455}, 1216 (2008)]. In order to observe fluctuations below the standard quantum limit, they utilized adiabatic passage to reduce the thermal noise to below that of thermal equilibrium at the minimum realizable temperature. We present a theoretical analysis that takes into account the spatial degrees of freedom of the system, allowing us to calculate the expected correlations at finite temperature in the system, and to verify the hypothesis of adiabatic passage by comparing the dynamics to the idealized model.
\end{abstract}

\submitto{\NJP}


\section{Introduction}

Interacting quantum systems are incredibly complex, allowing a range of states that span an enormous Hilbert space. Some of these quantum states, such as squeezed and entangled states, can be utilized for quantum information~\cite{Ekert1996,Gisin2002} and precision measurement~\cite{Wineland1994} purposes, potentially providing practical improvements over classical techniques. For instance, many common measurements based on quantum mechanical principles are usually limited to a much lower precision than allowed by the fundamental Heisenberg limit. The so-called standard quantum limit can be obtained by using the quantum analogue of classical states, such as the coherent state of light~\cite{WallsMilburn}, to probe quantities ranging from positions and velocities, to magnetic fields and atom numbers. One can improve the precision by using entangled or squeezed states as inputs to interferometric measurement procedures~\cite{Wineland1994,Bollinger1996}.

Experimentally realizing such quantum states presents many technical challenges. Macroscopic superpositions (or Schr\"odinger cat states) can in principle be used to perform Heisenberg-limited measurements~\cite{Bollinger1996}, but in practice are extremely difficult to engineer. There has been much success in creating squeezed states of light~\cite{Braunstein2005} and the spin of atomic samples~\cite{Julsgaard2001} that can be used to make measurements more accurate than the standard quantum limit. Interactions between samples of photons and/or atoms can create these squeezed states, but excess technical and thermal noise place practical limitations on the amount of squeezing generated. In order for quantum information and precision measurement techniques based on squeezed states to be practical, these excess fluctuations must be reduced.

Several methods of creating relative number squeezing and entanglement in ultra-cold gases have been proposed, and some realized, such as four-wave mixing~\cite{Perrin2007,Perrin2008,Ferris2009} and the Kerr effect~\cite{Haine2009}, molecular dissociation~\cite{Greiner2005a,Savage2007}, Josephson junction analogies~\cite{PitaevskiiStringari,Esteve2008} and coupling with squeezed light~\cite{Olsen2008}. Recently, Est\`eve \emph{et al.} reported number difference squeezing and entanglement between multiple ultra-cold atomic clouds~\cite{Esteve2008}. The experiment involved loading repulsive rubidium-87 atoms into two or more potential wells at low temperatures. The repulsion between the atoms reduces atomic bunching, and at zero temperature causes the fluctuations in the population differences between the wells to be smaller than the `classical' binomial distribution. This experiment was the first to perform accurate atom counting \emph{within each well}, while simultaneously having access to phase correlations between the wells through expansion and interference measurements. Combining these procedures, they observed relative number squeezing that could be used in principle to perform measurements at a precision 3.8 dB better than the standard quantum limit.

One of the cornerstone features of this experiment is a technique employed to reduce the level of thermal fluctuations of the quantum state. Producing multiple condensates by evaporative cooling into multiple wells produced results close to the standard quantum limit. Est\`eve \emph{et al.} found improved results by first cooling the atoms in a single trap, before modifying the potential to split the condensate between several wells. They postulated that this improvement is due to an adiabatic following of the quasiparticles representing the (quantized) fluctuations. Briefly, the energy of the excitations causing fluctuations is higher in the single condensate than after the splitting; therefore, the number of quasiparticle excitations at a given temperature is lower prior to splitting. As the potential is modified, the system is driven out of thermal equilibrium while the number of quasiparticles remains approximately fixed, and the level of fluctuations becomes lower than what would result from a thermal state at the minimum realizable temperature. In \Ref{Esteve2008}, the authors test this hypothesis by comparing with a simple, two-mode model, with reasonable agreement between the results.

Here we present a critique of the two-mode model and the hypothesis of adiabatic passage. Going beyond the two mode model reveals regimes where two modes are sufficient, or when the approximation fails. Higher-order spatial modes contribute to the physics when the wells are not well separated, if the atom number is not large, or at higher energy or temperature scales. We employ the perturbative Bogoliubov method~\cite{PethickSmith} taking into account the full multi-mode structure of the system and find the two-mode description is quite accurate for well-separated clouds. We compare the assumption of adiabatic passage to a model of full rethermalization, where the system remains in thermal equilibrium as the clouds are separated while the entropy of the isolated quantum system is constant. As the entropy of the total system depends on contributions from each spatial mode, the full multi-mode statistics need to be calculated for this comparison. We find that only the adiabatic model produces the dramatic improvement in squeezing observed in the experiment. Finally, we perform truncated Wigner~\cite{Steel1998} simulations of the dynamics and find they agree well with the adiabatic model for a sufficiently large system. Thus we can conclude that thermalization between the quasiparticle modes in this system is slow, and adiabatic passage is a good model for experiments performed on these timescales.

\section{System model}

\begin{figure}[t]
\begin{center}
\includegraphics[width=0.52\columnwidth]{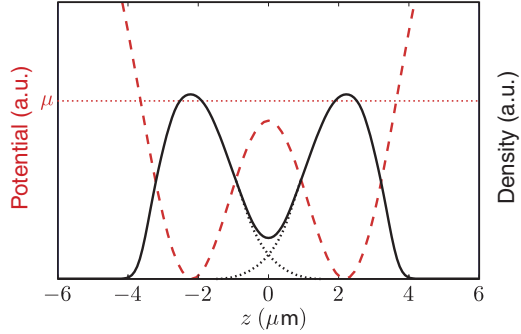}
\caption{A schematic of the double well potential (dashed red) and ground-state mean-field density (solid black) when the chemical potential $\mu$ is close to the barrier height. The ground state can be interpreted as a superposition of the atoms being in the left and right wells, sketched by the respective dotted-line ``wave-functions''. \label{fig_doublewell}}
\end{center}
\end{figure}

The experiment begins by evaporatively cooling a cloud of $^{87}$Rb atoms in an optical double well potential. The double-well potential is created by superimposing an optical lattice onto a harmonic dipole trap~\cite{Albiez2005,Esteve2008}. The potential in the region of the condensate is given by
\begin{equation}
  V(x,y,z) = \frac{1}{2}m\omega_{\perp}^2(x^2+y^2) + \frac{1}{2}m\omega_z^2 z^2 + \frac{V_L}{2}\cos(2k_L z),
\end{equation}
where the atomic mass of $^{87}$Rb is $m \approx 1.44 \times 10^{-25}$ kg. The trap is cylindrically symmetric and elongated in the direction of the lattice, $\omega_z < \omega_{\perp}$. The trap has radial trapping frequency $\omega_{\perp} \approx 2\pi \times 425$ Hz and, for the two-well experiment, a longitudinal trapping frequency of $\omega_z \approx 2\pi \times 60$ Hz. The optical lattice creates a barrier between two (or more) low energy regions, or wells, such as that depicted in \Fig{fig_doublewell} and is generated by interfering two phase-locked lasers at an angle producing a lattice spacing of $a \approx 5.7$ $\mu$m, where the lattice wave-vector is $k_L = \pi / a$. The peak-to-peak depth $V_L$ begins at $2\pi\hbar \times 430$ Hz before being ramped up.

The atoms are described by the Hamiltonian,
\begin{equation}
  \h{H} = \int \hd{\psi} \left[ \frac{-\hbar^2}{2m} \nabla^2 + V(\v{r},t) + \frac{U_0}{2}\hd{\psi}\h{\psi} \right] \h{\psi} \, d^3\v{r}. \label{doublewell_effective_hamiltonian}
\end{equation}
The interaction constant is $U_0 = 4\pi\hbar^2a_s/m$, where the s-wave scattering length is $a_s \approx 5.39$ nm.

We are interested in the low temperature limit, at or near the ground state of the system. To the lowest order of approximation, the solution is given by the ground-state mean-field wave-function $\psi_0$, which we find by 
integrating the imaginary-time Gross-Pitaevskii equation,
\begin{equation}
  -\hbar \frac{\partial \psi_0}{\partial \tau} = \frac{-\hbar^2}{2m}\nabla^2 \psi_0 + \bigl(V(\v{r}) + U_0|\psi_0|^2\bigr) \psi_0 - \mu \psi_0,
\end{equation}
until the state reaches equilibrium. The chemical potential $\mu$ is adjusted to give the required atom number $N_0 = \int |\psi_0|^2 \, d^3\v{r} = 1600$. 
If $\mu$ is close to the barrier height, the solution is that of two weakly-coupled condensates, sketched in \Fig{fig_doublewell}.

The peak-to-peak depth of the optical lattice begins at $2\pi\hbar \times 430$ Hz. As the gas is cooled, a Bose-Einstein condensate forms in the trap with the potential barrier between the wells significantly smaller than the chemical potential $\mu$ --- thus at this stage it can be considered a single condensate. The evaporative cooling method is unable to reduce the temperature $T$ significantly below the level of the chemical potential~\cite{Esteve2008}, so $k_B T \sim \mu$.

The potential barrier is slowly ramped up by increasing the intensity of the lasers. In the experiment, this was achieved adiabatically (i.e. slow enough that additional heating was not observed) at a rate where the peak-to-peak lattice potential $V_L$ increases by approximately $2\pi\hbar \times 2$ Hz per ms. The BEC is coherently split in two, whereupon the atoms configure themselves in a low energy state possessing relative number squeezing.

The experiment was then able to directly detect the atoms \emph{in situ}, and count the atoms in each well, $N_i$ to a greater accuracy than the shot-noise limit (i.e. $\Delta N_i < \sqrt{N_i}$). Defining the relative number squeezing, or normalized variance, as
\begin{equation}
   \xi = \Var\bigl[\h{N}_1 - \h{N}_2\bigr] / \langle \h{N}_1 + \h{N}_2 \rangle, \label{define_xi}
\end{equation}
their results indicate squeezing of up to -6.6 dB, or a reduction in the number difference variance by a factor of 4.5 compared to the binomial distribution expected from ideal, uncorrelated condensates. To quantatively determine the number variance we must go beyond the Gross-Pitaevskii description, which does not include either quantum or thermal fluctuations in the cloud.

\section{Bogoliubov analysis}

The Bogoliubov approach is a perturbative expansion valid in the weakly-interacting or large atom number limit~\cite{PethickSmith,Castin2001}. In either case, the ground state of the system is close to a coherent state given by the Gross-Pitaevskii equation. The Bogoliubov approach treats fluctuations about the mean-field perturbatively, by writing $\h{\psi}(\v{r},t) = [\psi_0(\v{r}) + \h{\delta\psi}(\v{r},t)]e^{-i\mu t}$,  and assuming $\h{\delta\psi}$ is `small' compared to $\psi_0$. The linearized solutions for $\h{\delta\psi}$ arising from \Eqn{doublewell_effective_hamiltonian} obey
\begin{equation}
  i\hbar \frac{\partial \h{\delta\psi}}{\partial t} = \left[\frac{-\hbar^2}{2m}\nabla^2 + V + 2U_0|\psi_0|^2 - \mu \right] \h{\delta\psi} + U_0 \psi_0^2 \hd{\delta\psi}, \\
\end{equation}
One can diagonalize the above linear equation (which couples $\h{\delta\psi}$ with $\hd{\delta\psi}$) into their respective eigenstates. The annihilation operators for the eigenstates have the form
\begin{equation}
 \h{b}_i = e^{i\varepsilon_i t/\hbar}\int u^{\ast}_i(\v{r}) \h{\delta\psi}(\v{r}) + v_i(\v{r}) \hd{\delta\psi}(\v{r}) d^3\v{r}.
\end{equation}
The eigenstates with eigenvalue $\epsilon_i$ are defined by the functions $u_i(\v{r})$ and $v_i(\v{r})$. Combining the above yields the Bogoliubov-de Gennes (BdG) equations
\begin{equation}
  \left[ \begin{array}{cc} \mathcal{L}_{\operatorname{GP}} + U_0 |\psi_0|^2 & -U_0 \psi_0^2 \\ U_0 \psi_0^{\ast 2} & -\mathcal{L}_{\operatorname{GP}} - U_0 |\psi_0|^2 \end{array} \right] \left[ \begin{array}{c} u_i \\ v_i \end{array} \right] = \epsilon_i \left[ \begin{array}{c} u_i \\ v_i \end{array} \right], \label{doublewell_BdG}
\end{equation}
where for brevity we have defined the Gross-Pitaevskii operator $\mathcal{L}_{\operatorname{GP}} = \frac{-\hbar^2}{2m}\nabla^2 + V(\v{r}) + U_0|\psi_0|^2 - \mu$. We implement the procedure presented in \Ref{Castin2001} to reliably solve \Eqn{doublewell_BdG}.

To efficiently solve the BdG equations, it is important to minimize the size of our three-dimensional spatial basis. Here we implement the harmonic oscillator basis~\cite{Blakie2008} as it effectively represents the trapped BEC in the lowest energy states while allowing efficient calculations of the matrix elements between the basis states $\phi_i(\v{r})$, for example $V_{ij} = \int \phi_i^{\ast}(\v{r}) \frac{V_L}{2}\cos(2k_L z) \phi_j(\v{r}) \, d^3\v{r}$ and $U_{ij} = \int |\psi_0^2(\v{r})| \phi^{\ast}_i(\v{r}) \phi_j(\v{r}) \, d^3\v{r}$ can be found accurately using the Gaussian quadrature sum rules~\cite{Blakie2005a,Blakie2008b}. We further optimize the calculations by taking advantage of $x$, $y$ and $z$ reflection symmetries, allowing us to reduce the computation cost by roughly a factor of $8^2$. 
The total system is truncated to the 2413 states having energy less than $90\hbar\omega_z$.

\begin{figure}[t]
\begin{center}
\includegraphics[width=0.67\columnwidth]{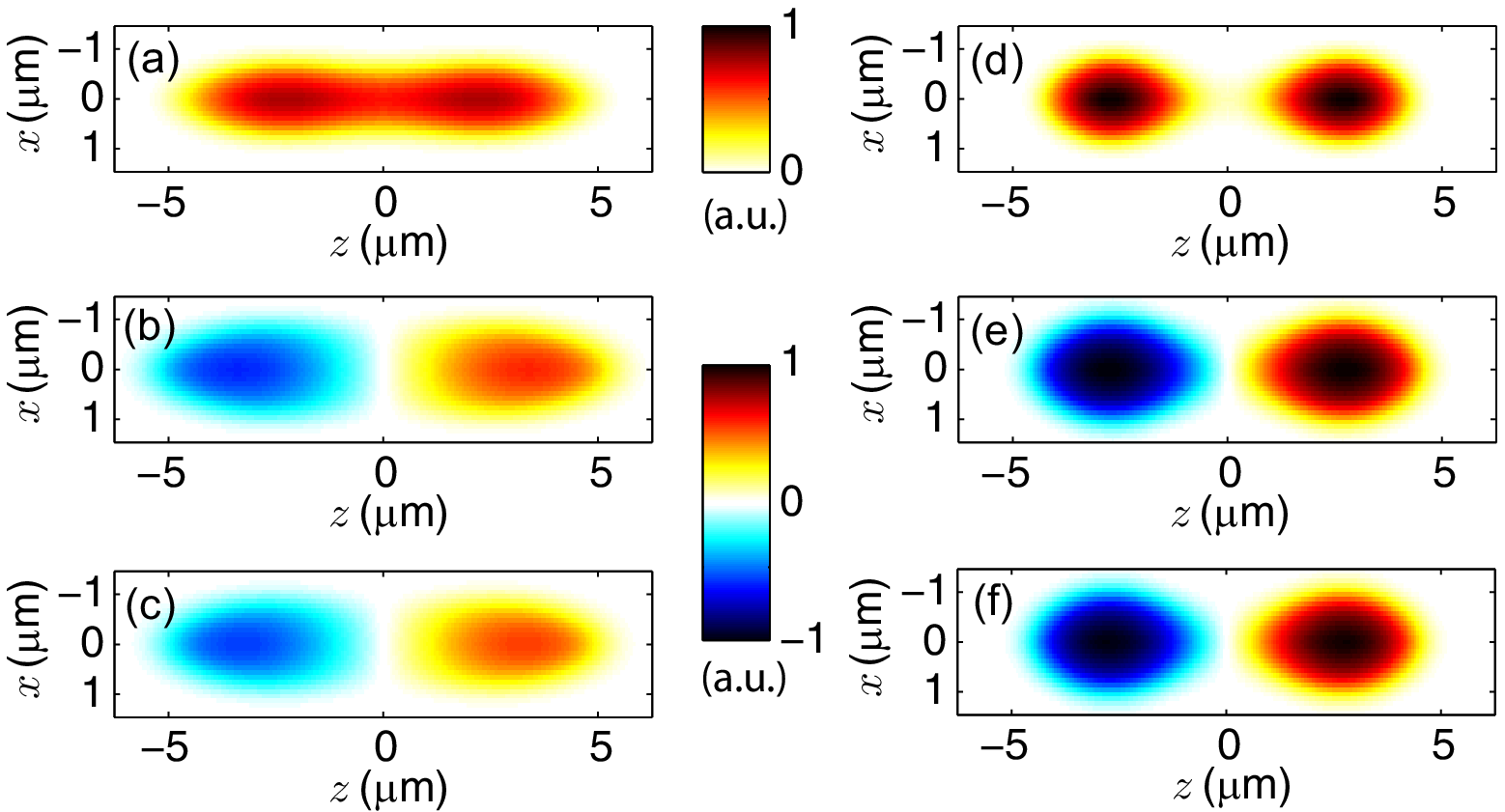}
\caption{Results of the Bogoliubov-de Gennes approach at a lattice depth of (a,b,c) 430 Hz and (d,e,f) 1650 Hz. (a,d) Density of the mean field $|\psi_0(\v{r})|^2$ through the $x$--$z$ plane. As the lattice is ramped the condensate splits into two well-defined wells. (b,c,e,f) The solutions for the lowest energy excitation, (b,d) $u_1(\v{r})$ and (c,e) $v_1(\v{r})$, also through the $x$-$z$ plane. This excitation removes particles from one well and transfers them to the other. \label{fig_density_u_v}}
\end{center}
\end{figure}

\begin{figure}[t]
\begin{center}
\includegraphics[width=0.38\columnwidth]{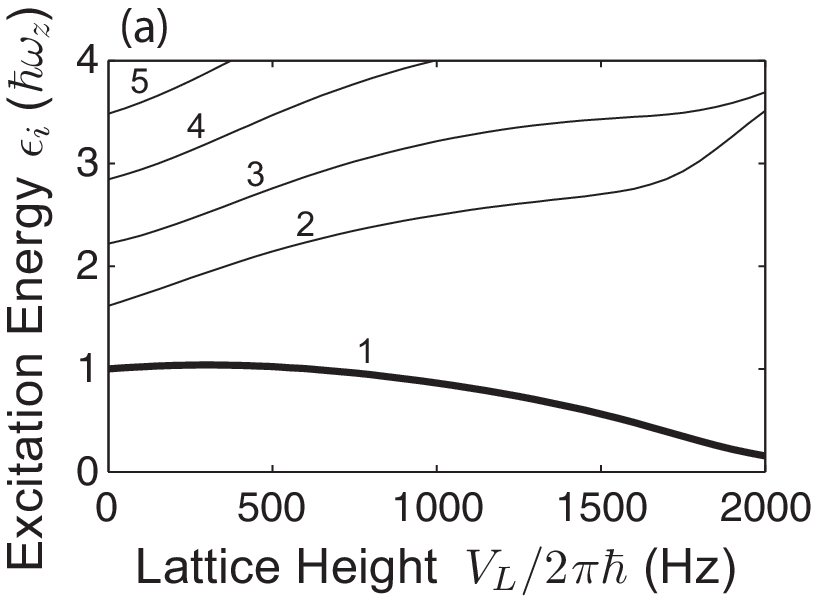} \hspace{1cm} \includegraphics[width=0.38\columnwidth]{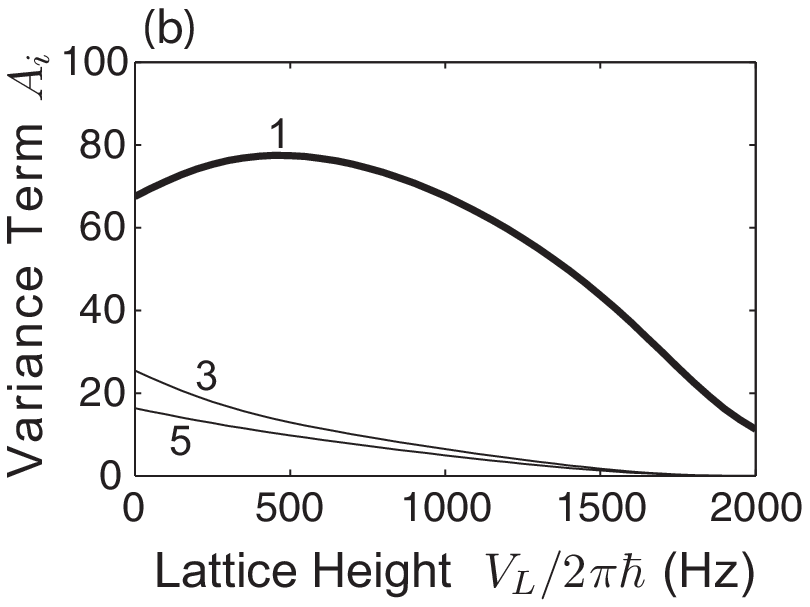}
\caption{(a) Energies of the lowest quasiparticle excitations as a function of lattice potential depth $V_L$. The lowest energy state is indicated and bold. (b) The first-order contributions to the number difference variance $A_i$, as given by equations~(\ref{variance_sum}) and (\ref{def_A}). The lowest energy state contributes most strongly to the number difference variance. \label{fig_doublewell_energies}}
\end{center}
\end{figure}

We display the condensate density and lowest energy quasiparticle mode solutions in \Fig{fig_density_u_v} for lattice depths of 430 Hz and 1650 Hz.  This lowest energy mode can be interpreted as transferring particles from one well to the other and corresponds to Josephson-style oscillations in population. For deep lattices the lowest energy mode has a significant overlap with the antisymmetrized ground state wave-function; that is $u(\v{r})$ and $v(\v{r})$ are approximately proportional to $\psi_0(\v{r}) \times P(z)$, where $P(z)$ is $-1$ for $z < 0$ and $+1$ otherwise (i.e. $P(z) = \Theta(z) - \Theta(-z)$, where $\Theta$ is the Heaviside step function). In \Fig{fig_doublewell_energies} we see that the energy of this lowest mode decreases as the lattice depth is increased, while the energy of the other modes increase. For $V_L \gtrsim 2\pi\hbar \times 1500$ Hz, the energy of the second excited mode is several times that of the lowest excitations, and at low temperatures one would expect the majority of quasiparticles in the lowest mode.

\subsection{Calculating correlations}

Based on the solutions to the Bogoliubov-de Gennes equations, one can calculate the variance of the atom number difference between the two wells as a function of the total atom number, temperatures and the potential. We begin by defining the regions of interest around each well, $\Omega_1$ and $\Omega_2$. For the double well we choose $\Omega_1$ to be the region with $z > 0$ and $\Omega_2$ to be the region with $z < 0$. The quantum operator describing the total number of atoms in the $i$th region is
 $ \h{N}_i = \int_{\Omega_i} \hd{\psi}(\v{r}) \h{\psi}(\v{r}) \, d^3\v{r} $.
Expanding the field operator about the mean field using the solutions to the Bogoliubov-de Gennes equations,
  $\h{\psi}(\v{r}) = \psi_0(\v{r}) + \sum_i \h{b}_i u_i(\v{r}) - \hd{b}_i v_i(\v{r})$,
and inserting into the above gives
\begin{eqnarray}
  \h{N}_i & = & \int_{\Omega_i} |\psi_0(\v{r})|^2 + \smash{\sum_j} \psi_0(\v{r}) \left(u_j^{\ast}(\v{r}) \hd{b}_j - v_j^{\ast}(\v{r}) \h{b}_j \right) \nonumber \\
  & & \hspace{2.32cm} + \; \psi_0^{\ast}(\v{r}) \left(u_j(\v{r}) \h{b}_j - v_j(\v{r}) \hd{b}_j \right) \nonumber \\
  & & + \vphantom{\Bigl(\Bigr)} \; \smash{\sum_{jk}} u_j^{\ast}(\v{r}) u_k(\v{r}) \hd{b}_j \h{b}_k - u_j^{\ast}(\v{r}) v_k(\v{r}) \hd{b}_j \hd{b}_k \nonumber \\
  & & \hspace{0.55cm} - \vphantom{\Bigl(\Bigr)}\; v_j^{\ast}(\v{r}) u_k(\v{r}) \h{b}_j \h{b}_k + v_j^{\ast}(\v{r}) v_k(\v{r}) \h{b}_j \hd{b}_k \; d^3\v{r}.
\end{eqnarray}

The system is symmetric under the transformation $z \rightarrow -z$, and so we expect the wells to be evenly balanced, $\langle \h{N}_1 - \h{N}_2 \rangle = 0$. We take an \emph{ansatz} for the many-body density matrix by assuming the Bogoliubov modes are in thermal (chaotic) states, with population $\langle \hd{b}_i \h{b}_i \rangle = [\exp(\epsilon_i/k_B T) - 1]^{-1}$. The variance of the number difference is therefore
\begin{eqnarray}
  \Var \bigl[ \h{N}_1 - \h{N}_2 \bigr] & = & N_0 \sum_j |A_j|^2 \left(\langle \hd{b}_j \h{b}_j \rangle + \langle \h{b}_j \hd{b}_j \rangle \right) \nonumber \\
  & & \hspace{-2cm}+ \sum_{i \ne j} \bigl(|B_{ji}|^2 + B^{\ast}_{ij}B_{ji}\bigr) \left( \langle \hd{b}_i \h{b}_i \rangle \langle \hd{b}_j \h{b}_j \rangle + \langle \h{b}_i \hd{b}_i \rangle \langle \h{b}_j \hd{b}_j \rangle \right) \nonumber \\
  & & \hspace{-1cm} + \sum_{i \ne j} \bigl(|C_{ij}|^2 + C_{ij}D_{ij}\bigr) \langle \hd{b}_i \h{b}_i \rangle \langle \h{b}_j \hd{b}_j \rangle \nonumber \\
  & & \hspace{-1cm}+ \sum_{i \ne j} \bigl(|D_{ij}|^2 + C_{ij}D_{ij}\bigr) \langle \h{b}_i \hd{b}_i \rangle \langle \hd{b}_j \h{b}_j \rangle, \label{variance_sum}
\end{eqnarray}
where we used the symmetry of the system to simplify the expression, and defined the following integrals for brevity:
\begin{eqnarray}
  A_j & = & \int P(z) \bigl( \psi_0^{\ast}(\v{r}) u_j(\v{r}) - \psi_0(\v{r}) v_j^{\ast}(\v{r}) \bigr) \, d^3\v{r} , \label{def_A} \\
  B_{ij} & = & \int P(z) u_i^{\ast}(\v{r}) v_j^{\ast}(\v{r}) \, d^3\v{r} , \\
  C_{ij} & = & \int P(z) u_i^{\ast}(\v{r}) u_j(\v{r}) \, d^3\v{r} , \\
  D_{ij} & = & \int P(z) v_i(\v{r}) v_j^{\ast}(\v{r}) \, d^3\v{r} ,
\end{eqnarray}
remembering $P(z)$ is $-1$ for $z < 0$ and $+1$ otherwise. We note that some the terms in equation (\ref{variance_sum}) go beyond the accuracy of the first-order Bogoliubov method used to derive them, and many authors therefore choose to disregard these terms. Here these terms are retained, without loss of accuracy, to facilitate a better comparison with the truncated Wigner simulations in section~\ref{sec_dyn_sim}, where the full density matrix resulting from thermally-occupied Bogoliubov modes is implemented as initial conditions.

For large $N_0$ and small fluctuations, the dominant contributions to the number variance come from the first sum in \Eqn{variance_sum}, and in particular the lowest energy (Josephson) mode has the largest contribution. In figure~\ref{fig_doublewell_energies}~(b), we see that for deep lattices $|A_1|^2$ is much larger than those corresponding to other modes, and simultaneously the quasiparticle population in the lowest energy mode is expected to be greatest. Together these might justify the two mode approximation for lattice intensities $V_L \gtrsim 2\pi\hbar \times 1500$ Hz, whereas higher energy excitations are important for weaker lattices.

This would indeed be the case for systems of larger particle number. However, in this system the additional contributions to \Eqn{variance_sum} that a two mode model would fail to capture are significant. Even for well separated wells ($V_L \gtrsim 2\pi\hbar \times 1500$ Hz) these extra terms contribute approximately one third of the total variance at zero temperature. In this regime, we expect a small loss of precision in our calculations due to the fact that those terms proportional to $B_{ij}$, $C_{ij}$ and $D_{ik}$ go beyond the accuracy of the first-order Bogoliubov analysis used, and may be poorly approximated.

\begin{figure}[t]
\begin{center}
\includegraphics[width=0.48\columnwidth]{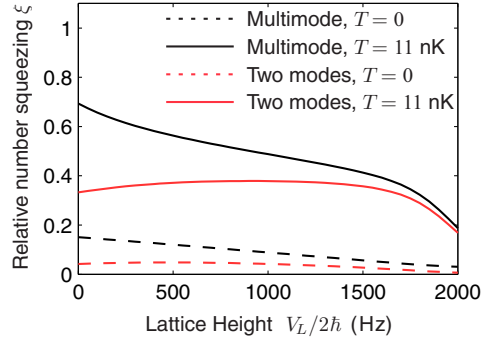}
\caption{The number difference squeezing $\xi$ [equation~(\ref{define_xi})] as a function of barrier height for the full multi-mode analysis (black lines) and a simplified, two-mode model involving just the lowest Bogoliubov excitation (red lines). The dashed lines are the results at zero temperature, while the solid lines are for $T = 11$ nK. The models agree best for large lattice heights, or well-separated clouds. \label{fig_two_mode_comparison}}
\end{center}
\end{figure}

In figure~\ref{fig_two_mode_comparison} we compare the results of the full multi-mode analysis presented here, and a simplified model including just the lowest Bogoliubov excitation. This simplified model is somewhat akin to a two-mode approximation, and we see that the two models show reasonable agreement for large barrier heights. In this regime, the two atomic clouds are well-separated and the two-mode approximation is expected to be accurate.

\begin{figure}[t]
\begin{center}
\includegraphics[width=0.48\columnwidth]{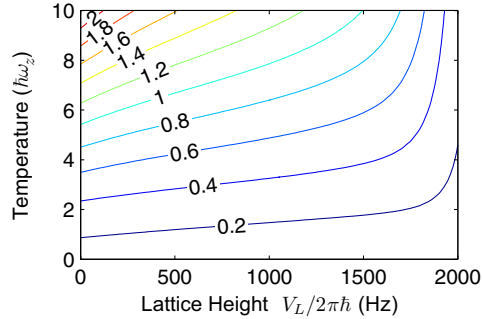}
\caption{A contour plot of the number difference squeezing $\xi$ as a function of temperature and barrier height in the double well system. Low temperatures and large barriers produce the least relative number fluctuations. \label{fig_contour}}
\end{center}
\end{figure}

In figure \ref{fig_contour} we plot the number squeezing $\xi$ [equation~(\ref{define_xi})] as a function of barrier height and temperature. Clearly, best squeezing is obtained at lower temperatures and with larger barrier heights.

\subsection{Adiabatic Passage}

We now compare our results to the double-well experiment performed at Universit\"at Heidelberg and examine the hypothesis of adiabatic passage presented in~\Ref{Esteve2008}. The experiment found improved number squeezing when slowly raising the barrier after evaporation compared to performing evaporative cooling with the barrier at fixed height. If the lattice ramp is slow compared to the energy gaps in the Bogoliubov spectrum, $\epsilon_1$, $\epsilon_2 - \epsilon_1$, \emph{et cetera}, one might expect the population in each quasiparticle mode to be fixed during the evolution. If this were true, then the system would no longer be in thermal equilibrium, as the ratio of the excitation energies changes during the ramp (see \Fig{fig_doublewell_energies}). The ramp ends with fewer quasiparticle excitations in the lowest mode than one would expect at realizable temperatures.

However, for sufficiently slow ramps one might expect the system to rethermalize, so that the population in each Bogoliubov changes to conform to a new global temperature as the energy levels move, and we also consider this possibility. The evolution of the potential, and therefore the quasiparticle energies, is slow compared to the gaps in the Bogoliubov spectrum, but it is unclear whether the evolution is adiabatic with respect to the full many-body Hamiltonian. 
The Bogoliubov description ignores interactions between the quasiparticles that can lead to a redistribution of excitations and a return to thermal equilibrium.

An isolated system undergoing quasi-static evolution will have constant entropy. The entropy of mode $i$ in a thermal state, in terms of the mean occupation, is~\cite{Huang}
\begin{equation}
  S_i = \langle \hd{b}_i \h{b}_i \rangle \ln \left(1 + \frac{1}{\langle \hd{b}_i \h{b}_i \rangle}\right) + \ln\left(1 + \langle \hd{b}_i \h{b}_i \rangle \right).
\end{equation}
The entropy of the total system is $S = \sum_i S_i$. As the lattice height, and therefore excitation spectrum, varries, we calculate the new temperature required to keep the system in thermal equilibrium with fixed entropy. As many modes make significant contributions to the total entropy, and entropy is exchanged between modes, this analysis is not possible with a two-mode model.

\begin{figure}[t]
\begin{center}
\includegraphics[width=0.48\columnwidth]{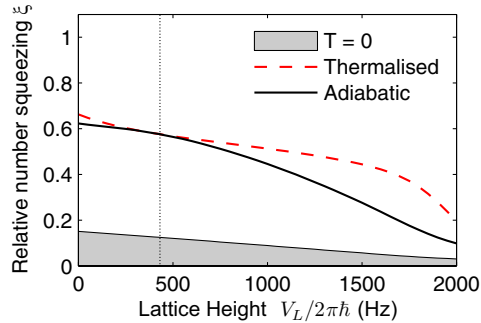}
\caption{A comparison of different thermalization models for the relative number squeezing as the lattice height is slowly ramped from 430 Hz to 1650 Hz (indicated by the vertical dotted lines). In all cases, at a lattice depth of 430 Hz we take a thermal distribution with $T = \mu/6$. The dashed red line is the squeezing where we assume the state remains in thermal equilibrium with constant entropy. The solid black line is for the adiabatic model where the number of quasiparticle excitations in each mode does not change. The adiabatic model agrees with the experimental observation of significantly decreased number fluctuations when slowly ramping up the lattice. \label{fig_doublewell_compare}}
\end{center}
\end{figure}

We compare the two models for thermalization as the lattice potential is increased in~\Fig{fig_doublewell_compare}. In all cases we begin in thermal equilibrium with a lattice strength of 430 Hz at a temperature of $k_B T \approx \mu / 6$, chosen to roughly correspond to the experimentally observed fluctuations.

As can be seen from \Fig{fig_doublewell_compare}, the adiabatic process predicts a level of fluctuations significantly lower than the fully thermalizing model. In a realistic experiment one might expect the entropy to increase slightly, resulting in even greater fluctuations. As the experiment observes strongly improved number difference squeezing as the lattice is ramped, one can rule out the possibility of rethermalization on the timescales of the experiment. On the other hand, the assumption of adiabatic passage agrees well with the experimental results, and we conclude that the mechanism of effective cooling proposed in~\cite{Esteve2008} is correct.

\section{Dynamical simulation}

\label{sec_dyn_sim}
In this section we perform a direct dynamical simulation using the truncated Wigner~\cite{Steel1998,Ferris2008,Ferris2009,Blakie2008} method and compare the results to perfect adiabatic passage. As pointed out in the previous section, the number fluctuations of the double-well system realized by~\cite{Esteve2008} are sensitive to second-order interactions between the linearized Bogoliubov modes. For the experimental parameters, the truncated Wigner simulations show dynamics that is inconsistent with the linearized Bogoliubov-de Gennes solutions -- in particular, the quasiparticle populations and number difference variance are not constant in time even for a fixed Hamiltonian. The Bogoliubov fluctuations are too large to be considered perturbative, and the linearized approximation breaks down.

Nevertheless, we can probe adiabatic passage dynamically by investigating a slightly different system where the perturbations $\h{\delta \psi}$ are small compared to $\psi_0$, and linearization is accurate. To minimize the changes to the system we investigate a system of more atoms with smaller interactions, such that the product $N_0U_0$ is unchanged. In principle, such a change could be realized with a Feshbach resonance~\cite{PethickSmith}, although this would be challenging.

The truncated Wigner method is an approximate phase space method that takes advantage of the Wigner representation of a quantum field. Truncating the higher-order derivatives from the equation of motion for the Wigner function results in a classical Liouville equation that can be efficiently sampled stochastically. Thus we reduced the problem of solving a non-linear equation for a quantum field $\h{\psi}(\v{r})$ to that of the non-linear motion of an ensemble of classical fields $\psi(\v{r})$. The resulting trajectories obey the (real-time) Gross-Pitaevskii equation, where the initial conditions are sampled from the initial Wigner distribution. Expectation values of symmetrically ordered operators are equal to the ensemble average of their classical counterparts; however care must be taken when using a finite, non-uniform basis~\cite{Blakie2008}.

The system we simulate has $1.6 \times 10^5$ atoms and a scattering length $a_s = 5.39 \times 10^{-11}$ m, and begins in thermal equilibrium at $V_L = 2\pi\hbar \times 430$ Hz with a temperature $T \approx \mu / 6$. This system has 100 times more atoms than the experiment, but identical mean-field and Bogoliubov-de Gennes solutions. The initial state is populated by a coherent state condensate surrounded by Bogoliubov modes in thermal (chaotic) states with populations given by the Bose-Einstein statistics~\cite{PethickSmith}. The Wigner distribution for such a state is Gaussian and therefore straightforward to randomly sample. We evolve 4000 independent trajectories with the Gross-Pitaevskii equation, while the lattice is ramped up at a rate $2\pi\hbar\times2$ Hz/ms. As we use the non-uniform Hermite-Gauss basis for our simulation, care must be taken to extract the correct quantum expectation values from the averaged data. For arbitrary basis $\{\phi_i(\v{r})\}$ such that $\psi(\v{r}) = \sum c_i \phi_i(\v{r})$ and matrix $P$ defined by
\begin{equation}
 P_{ij} = \int \phi_i^{\ast}(\v{r}) P(\v{r}) \phi_j(\v{r}) \; d^3\v{r},
\end{equation}
we find the number difference
\begin{equation}
 \langle \h{N}_1 - \h{N}_2 \rangle = \int \hd{\psi}(\v{r}) P(\v{r}) \h{\psi}(\v{r}) = \overline{\v{c}^{\dag} P \v{c}} - \operatorname{Trace}\bigl[P\bigr]/2,
\end{equation}
and the number difference squared
\begin{equation}
 \langle \bigl( \h{N}_1 - \h{N}_2 \bigr)^2 \rangle = \overline{\left(\v{c}^{\dag} P \v{c} - \operatorname{Trace}\bigl[P\bigr]/2 \right)^2} - \operatorname{Trace}\bigl[P^2\bigr]/4.
\end{equation}
These relations allow us to efficiently calculate the number difference variance without transforming to the spatial basis, while simplifying the symmetric Wigner corrections for a non-uniform basis~\cite{Blakie2008}.

\begin{figure}[t]
\begin{center}
\includegraphics[width=0.48\columnwidth]{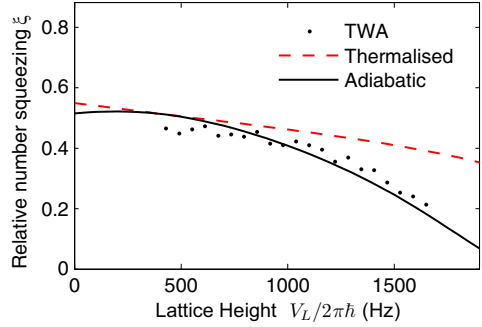}
\caption{A comparison of the truncated Wigner dynamics (points), a model of complete adiabatic following (black solid line), and the fully thermalized model (red dashed line) for the system with $1.6 \times 10^5$ atoms and a small scattering length. The dynamics agree with the model of adiabatic passage. \label{fig_wigner}}
\end{center}
\end{figure}

Our dynamical results are plotted in \Fig{fig_wigner}, along with the predictions for perfect adiabatic following and full thermalization obtained by the same procedure as in the previous section. We see that the number difference variance agrees with adiabatic model, validating the hypothesis of adiabatic following on this timescale.

\section{Systems with different geometry}

It is worth discussing changes that occur in multi-well systems of different geometry. First, double well traps can be created in a variety of ways, such as the fully optical approach analysed here~\cite{Esteve2008} or with magnetic traps on atom chips~\cite{Schumm2005}. The latter setup results in two parallel, elongated condensates (with $\omega_x \ll \omega_y$ and $\omega_z$). In such a system, one would expect long wavelength excitations along the transverse dimension $x$, some of which will have energy less than the Josephson mode. Similarly, one expects multiple Josephson-type modes within a small energy band, coupling the typical double-well excitation with low lying transverse modes (the limit of two infinite condensates can be found in \cite{Tommasini2003}). In this case, a multimode description such as is presented here is essential to analysing these systems. It is unclear, however, if such a geometry is advantageous for creating sub-shot noise number correlations. The increased density of states about the Josephson mode may make adiabatic following more difficult. Furthermore, a clear definition of the relative phase of the condensates, as investigated in reference~\cite{Esteve2008}, may be problematic as the phase may vary along the long axis of the system.

In reference~\cite{Esteve2008} the authors also investigated systems of multiple wells. They presented improved results for the inner pair of six coupled wells, which was theoretically investigated using a six-mode Bogoliubov theory. We performed a multi-mode analysis of this system, which involved an increased computational difficulty. In the multi-mode picture, we observed a band of five Josephson-type states of decreasing energy as the barrier height is increased --- one mode for each inter-well coupling --- while the remaining excitations had increasing energy. Thus, we can conclude that the adiabatic following technique should also be benificial in this arrangement, as was indeed verified experimentally~\cite{Esteve2008}. We were unfortunately unable to repeat the full analysis for that system because of the increased numberical difficulty.

\section{Conclusion}

We have performed a multi-mode Bogoliubov analysis of coupled Bose-Einstein condensate systems to find the low temperature physics and calculate the population fluctuations between the two wells. At very low temperatures, repulsive interactions induce relative number squeezing and even entanglement between the wells. We present a critical analysis of the adiabatic passage hypothesis employed in~\Ref{Esteve2008} as the optical lattice separating the wells is ramped up. Our results indicate that adiabatic evolution of the quasiparticles is possible, causing a decrease in number fluctuations as the lattice height is increased. We conclude that the isolated system can be driven out of thermal equilibrium, to reduce fluctuations to below what is possible at thermal equilibrium. It remains to be seen if this procedure can be improved to reach the quantum regime, where the vacuum fluctuations dominate the physics.

There are several limitations to the Bogoliubov approximation; we are unable to investigate either the low atom number or high temperature limits. The transition between anti-bunching due to repulsion and boson bunching at higher temperatures could be investigated by non-perturbative methods. Quantum Monte Carlo techniques~\cite{Ceperley1986} provide the most accurate and reliable results for high temperature bosonic systems. The Positive-P phase space method can be implemented to find thermal statistics of a Bose gas~\cite{Drummond2004}. The Popov approximation~\cite{PethickSmith} is similar to the approach used here, but considers the interactions between the excited modes and back-action on the condensate, and may be an appropriate method for exploring this regime. Related to these, Sinatra \emph{et al.}~\cite{Sinatra2000} have proposed a method for sampling the statistics of a Bose gas described by a Gaussian density matrix.

\ack

We would like to thank Markus Oberthaler, Jerome Est\`eve, Stefano Giovanazzi, Jason McKeever and Joseph Thywissen for stimulating discussions. This research was funded by the Australian Research Council Centre of Excellence for Quantum-Atom Optics.

\section*{References}

\bibliography{andy}

\begin{thebibliography}{10}

\bibitem{Ekert1996}
A~Ekert and R~Jozsa.
\newblock Quantum computation and {S}hor's factoring algorithm.
\newblock {\em Rev. Mod. Phys.}, {68}({3}):733--753, {JUL} {1996}.

\bibitem{Gisin2002}
Nicolas Gisin, Gr{\'e}goire Ribordy, Wolfgang Tittel, and Hugo Zbinden.
\newblock Quantum cryptography.
\newblock {\em Rev. Mod. Phys.}, 74:145, 2002.

\bibitem{Wineland1994}
D.~J. Wineland, J.~J. Bollinger, W.~M. Itano, and D.~J. Heinzen.
\newblock Squeezed atomic states and projection noise in spectroscopy.
\newblock {\em Phys. Rev. A}, 50:67, 1994.

\bibitem{WallsMilburn}
D.~F. Walls and Gerard~J. Milburn.
\newblock {\em Quantum {O}ptics}.
\newblock Springer, Berlin, second edition, 2006.

\bibitem{Bollinger1996}
J.~J~. Bollinger, Wayne~M. Itano, D.~J. Wineland, and D.~J. Heinzen.
\newblock Optimal frequency measurements with maximally correlated states.
\newblock {\em Phys. Rev. A}, 54(6):4649(R), Dec 1996.

\bibitem{Braunstein2005}
S.~L. Braunstein and P.~van Loock.
\newblock Quantum information with continuous variables.
\newblock {\em Rev. Mod. Phys.}, 77(2):513--577, April 2005.

\bibitem{Julsgaard2001}
Brian Julsgaard, Alexander Kozhekin, and Eugene~S. Polzik.
\newblock Experimental long-lived entanglement of two macroscopic objects.
\newblock {\em Nature}, 413:400, 2001.

\bibitem{Perrin2007}
A.~Perrin, H.~Chang, V.~Krachmalnicoff, M.~Schellekens, D.~Boiron, A.~Aspect,
  and C.~I. Westbrook.
\newblock Observation of atom pairs in spontaneous four-wave mixing of two
  colliding {B}ose-{E}instein condensates.
\newblock {\em Phys. Rev. Lett.}, 99(15):150405, October 2007.

\bibitem{Perrin2008}
A.~Perrin, C.~M. Savage, D.~Boiron, V.~Krachmalnicoff, C.~I. Westbrook, and
  K.~V. Kheruntsyan.
\newblock Atomic four-wave mixing via condensate collisions.
\newblock {\em New J. Phys.}, 10:045021, 2008.

\bibitem{Ferris2009}
Andrew~J. Ferris, Murray~K. Olsen, and Matthew~J. Davis.
\newblock Atomic entanglement generation and detection via degenerate
  four-wave-mixing of a {B}ose-{E}instein condensate in an optical lattice.
\newblock {\em Phys. Rev. A}, 79:043634, 2009.

\bibitem{Haine2009}
S.~A. Haine and M.~T. Johnsson.
\newblock Dynamic scheme for generating number squeezing in {B}ose-{E}instein
  condensates through nonlinear interactions.
\newblock {\em Phys. Rev. A}, 80:023611, 2009.

\bibitem{Greiner2005a}
M.~Greiner, C.~A. Regal, J.~T. Stewart, and D.~S. Jin.
\newblock Probing pair-correlated fermionic atoms through correlations in atom
  shot noise.
\newblock {\em Phys. Rev. Lett.}, 94:110401, 2005.

\bibitem{Savage2007}
C.~M. Savage and K.~V. Kheruntsyan.
\newblock Spatial pair correlations of atoms in molecular dissociation.
\newblock {\em Phys. Rev. Lett.}, 99:220404, 2007.

\bibitem{PitaevskiiStringari}
Lev Pitaevskii and Sandro Stringari.
\newblock {\em Bose-{E}instein Condensation}.
\newblock Oxford University Press, 2003.

\bibitem{Esteve2008}
J.~Est{\`e}ve, C.~Gross, A.~Weller, S.~Giovanazzi, and M.~K. Oberthaler.
\newblock Squeezing and entanglement in a {B}ose-{E}instein condensate.
\newblock {\em Nature}, 455:1216, October 2008.

\bibitem{Olsen2008}
M.~K. Olsen, S.~A. Haine, A.~S. Bradley, and J.~J. Hope.
\newblock From squeezed atom lasers to teleportation of massive particles.
\newblock {\em European Physical Journal-Special Topics}, 160:331--342, July
  2008.

\bibitem{PethickSmith}
C.~J. Pethick and H.~Smith.
\newblock {\em Bose-{E}instein Condensation in Dilute Gases}.
\newblock Cambridge University Press, 2001.

\bibitem{Steel1998}
M.~J. Steel, M.~K. Olsen, L.~I. Plimak, P.~D. Drummond, S.~M. Tan, M.~J.
  Collett, D.~F. Walls, and R.~Graham.
\newblock Dynamical quantum noise in trapped {B}ose-{E}instein condensates.
\newblock {\em Phys. Rev. A}, 58:4824, 1998.

\bibitem{Albiez2005}
M.~Albiez, R.~Gati, J.~F{\"o}lling, S.~Hunsmann, M.~Cristiani, and M.~K.
  Oberthaler.
\newblock Direct {O}bservation of {T}unneling and {N}onlinear {S}elf-{T}rapping
  in a {S}ingle {B}osonic {J}osephson {J}unction.
\newblock {\em Phys. Rev. Lett.}, 95:010402, 2005.

\bibitem{Castin2001}
Yvan Castin.
\newblock {\em `Coherent atomic matter waves', Lecture Notes of Les Houches
  Summer School 1999}, pages 1--136.
\newblock EDP Sciences and Springer-Verlag, 2001.

\bibitem{Blakie2008}
P.~B. Blakie, A.~S. Bradley, M.~J. Davis, R.~J. Ballagh, and C.~W. Gardiner.
\newblock Dynamics and statistical mechanics of ultra-cold {B}ose gases using
  c-field techniques.
\newblock {\em Advances in Physics}, 57:363, 2008.

\bibitem{Blakie2005a}
P.~B. Blakie and M.~J. Davis.
\newblock {P}rojected {G}ross-{P}itaevskii equation for harmonically confined
  {B}ose gases at finite temperature.
\newblock {\em Phys. Rev. A}, 72:063608, 2005.

\bibitem{Blakie2008b}
P.~Blair Blakie.
\newblock Numerical method for evolving the projected gross-pitaevskii
  equation.
\newblock {\em Phys. Rev. E}, 78:026704, 2008.

\bibitem{Huang}
Kerson Huang.
\newblock {\em Statistical Mechanics}.
\newblock John Wiley and Sons, second edition, 1988.

\bibitem{Ferris2008}
Andrew~J. Ferris, Matthew~J. Davis, Reece~W. Geursen, P.~Blair Blakie, and
  Andrew~C. Wilson.
\newblock Dynamical instabilities of {B}ose-{E}instein condensates at the band
  edge in one-dimensional optical lattices.
\newblock {\em Phys. Rev. A}, 77:012712, 2008.

\bibitem{Schumm2005}
T.~Schumm, S.~Hofferberth, L.~M. Andersson, S.~Wildermuth, S.~Groth,
  I.~Bar-Joseph, J.~Schmiedmayer, and P.~{Kr\"uger}.
\newblock Matter-wave interferometry in a double well on an atom chip.
\newblock {\em Nature Physics}, 1:57, 2005.

\bibitem{Tommasini2003}
Paolo Tommasini, E.~J.~V. {de Passos}, A.~F.~R. {de Toledo Piza}, M.~S.
  Hussein, and E.~Timmermans.
\newblock Bogoluibov theory for mutually coherent condensates.
\newblock {\em Phys. Rev. A}, 67:023606, 2003.

\bibitem{Ceperley1986}
D.~Ceperley and B.~Alder.
\newblock Quantum {M}onte-{C}arlo.
\newblock {\em Science}, 231:555, 1986.

\bibitem{Drummond2004}
P.~D. Drummond, P.~Deuar, and K.~V. Kheruntsyan.
\newblock Canonical {B}ose gas simulations with stochastic gauges.
\newblock {\em Phys. Rev. Lett.}, 92:40405, 2004.

\bibitem{Sinatra2000}
A.~Sinatra, Y.~Castin, and C.~Lobo.
\newblock A {M}onte {C}arlo formulation of the {B}ogolubov theory.
\newblock {\em J. Mod. Opt.}, 47:2629, 2000.

\end{thebibliography}
\bibliographystyle{unsrt}

\end{document}